# Deep Learning Approach for the Diagnosis of Pediatric Pneumonia Using Chest X-ray Imaging

Fatemeh Hosseinabadi[1], Mohammad Mojtaba Rohani[2]

[1] Assistant Professor of Radiology, Zahedan University of medical Sciences, Iran

[2] Department of Radiology, Poursina Hospital, Guilan University of Medical Sciences, Rasht, Iran

**Abstract**

Pediatric pneumonia remains a leading cause of morbidity and mortality in children worldwide. Timely and accurate diagnosis is critical but often challenged by limited radiological expertise and the physiological and procedural complexity of pediatric imaging. This study investigates the performance of state-of-the-art convolutional neural network (CNN) architectures—ResNetRS, RegNet, and EfficientNetV2—using transfer learning for the automated classification of pediatric chest X-ray images as either pneumonia or normal.A curated subset of 1,000 chest X-ray images was extracted from a publicly available dataset originally comprising 5,856 pediatric images. All images were preprocessed and labeled for binary classification. Each model was fine-tuned using pretrained ImageNet weights and evaluated based on accuracy and sensitivity. RegNet achieved the highest classification performance with an accuracy of 92.4% and a sensitivity of 90.1%, followed by ResNetRS (accuracy: 91.9%, sensitivity: 89.3%) and EfficientNetV2 (accuracy: 88.5%, sensitivity: 88.1%).

## Introduction

Pediatric pneumonia remains one of the most critical health challenges affecting children globally, accounting for over 700,000 deaths annually in children under the age of five, according to the World Health Organization. It is caused by a variety of pathogens, including bacteria such as Streptococcus pneumoniae, viruses like respiratory syncytial virus (RSV), and, less commonly, fungi. The clinical presentation of pneumonia in children often includes fever, cough, rapid breathing, and chest indrawing, but these symptoms can overlap with other common pediatric illnesses such as asthma or bronchiolitis, complicating clinical diagnosis. Additionally, young children may not be able to effectively communicate their symptoms, making objective diagnostic tools all the more essential. Prompt diagnosis is vital not only to initiate timely treatment but also to avoid unnecessary antibiotic usage and prevent disease progression, hospitalization, or death [1].

Medical imaging, particularly chest X-ray radiography, is the most widely used diagnostic tool for confirming suspected cases of pneumonia in children [2-4]. It offers a non-invasive means of visualizing lung infiltrates, consolidation, and pleural effusion—hallmarks of pneumonia. However, interpreting pediatric chest X-rays poses unique challenges. Children have anatomical differences compared to adults, such as smaller thoracic cavities and less-defined lung structures, which can complicate image interpretation. Moreover, acquiring high-quality images can be technically difficult due to patient movement, lack of cooperation, or suboptimal exposure settings. In resource-limited settings, the situation is further exacerbated by a shortage of trained radiologists or pediatric imaging specialists, leading to diagnostic delays or inaccuracies. Consequently, there is a growing demand for automated and reliable diagnostic support systems.

In recent years, machine learning (ML) and deep learning (DL) technologies have shown substantial promise in addressing such diagnostic challenges. Machine learning encompasses a broad set of algorithms that learn from data to make predictions or decisions without being explicitly programmed [5-8]. Deep learning, a subfield of ML, involves neural networks with multiple layers that automatically extract complex features from raw data. Among various DL architectures, convolutional neural networks (CNNs) have gained prominence for their exceptional performance in image-related tasks. CNNs are capable of learning hierarchical patterns, from simple edges and textures to more abstract structures, making them particularly suitable for medical image analysis. These models not only reduce the burden on clinicians but also introduce consistency and objectivity to the diagnostic process.

The application of deep learning in medicine has rapidly expanded, demonstrating state-of-the-art results across a range of clinical domains. In radiology, CNN-based models have been developed to detect diabetic retinopathy in retinal images, classify skin lesions in dermatology, and identify lung nodules or tuberculosis in chest X-rays. These systems can assist in triage, serve as second readers, or even function as autonomous diagnostic tools in environments lacking human expertise. In addition to classification, deep learning has been used for medical image segmentation, anomaly detection, and survival prediction [9-12]. The integration of AI in healthcare promises to improve diagnostic speed, accuracy, and accessibility, particularly in underserved or high-volume clinical settings. Building on these advancements, this study explores the use of three cutting-edge CNN architectures—ResNetRS, RegNet, and EfficientNetV2—for the detection of pediatric pneumonia using chest X-ray images, with the goal of improving diagnostic outcomes through automated image classification.

In this study, we present a comparative analysis of three modern convolutional neural network architectures—ResNetRS, RegNet, and EfficientNetV2—for the task of pediatric pneumonia classification using chest X-ray images. We create a balanced subset of 1,000 images from a publicly available pediatric chest radiograph dataset and apply standardized preprocessing and augmentation techniques to improve model robustness. Leveraging transfer learning with pretrained ImageNet weights, each model is fine-tuned and evaluated based on its classification accuracy and sensitivity.

**Related Work**

The use of deep learning in medical imaging has garnered significant attention over the past decade, with numerous studies demonstrating its efficacy in diagnosing thoracic diseases, including pneumonia. One of the most influential works in this domain is by [1], which introduced a large, publicly available pediatric chest X-ray dataset comprising over 5,000 images labeled as pneumonia (bacterial or viral) or normal. Using a transfer learning approach with the InceptionV3 architecture, their model achieved over 90% accuracy in distinguishing pneumonia from normal cases, showcasing the potential of deep learning in pediatric diagnostic imaging.

Following this foundational work, [13] developed CheXNet, a 121-layer DenseNet trained on the NIH ChestX-ray14 dataset and achieved radiologist-level performance in detecting pneumonia in adult chest X-rays. While not specifically targeted at pediatric patients, this work laid the groundwork for the use of deeper architectures in chest pathology detection. Building upon these advances, researchers have sought to adapt and evaluate CNNs specifically for pediatric populations, recognizing the anatomical and pathological differences.

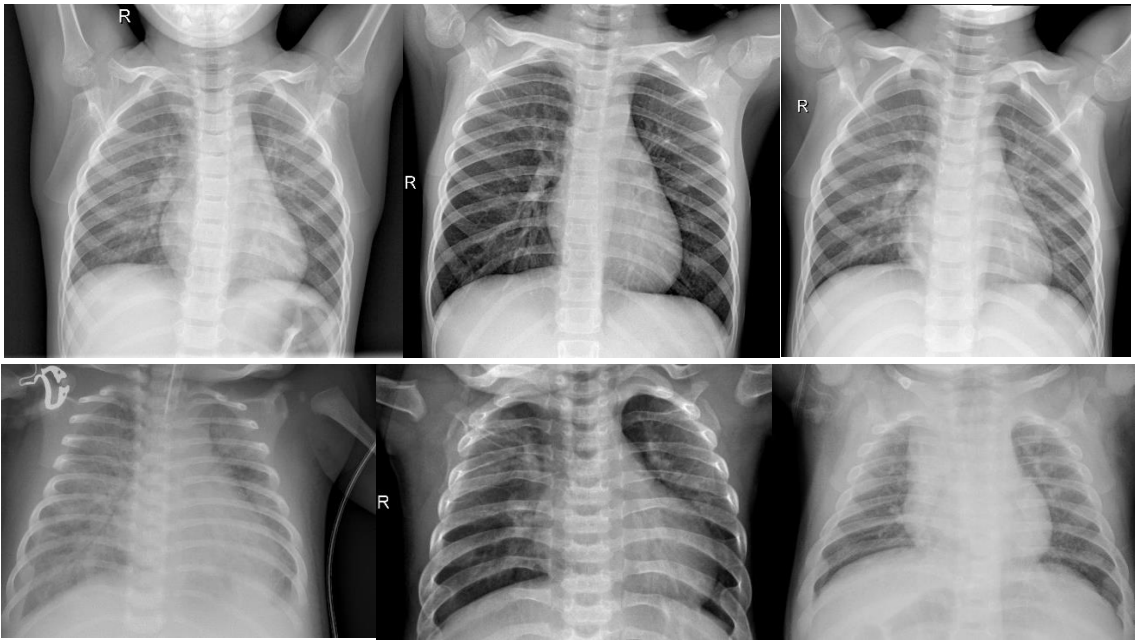

*Figure 1 Examples of Pediatric Chest X-ray. Top: Normal, Down: Pneumonia [1]*

Researchers in [14] explored the performance of VGG16 and ResNet50 architectures in classifying pediatric chest X-rays and emphasized the importance of dataset balance and augmentation in achieving high sensitivity. Similarly, [15] applied a deep convolutional neural network for pneumonia diagnosis and highlighted the potential of automated systems in reducing diagnostic workload. More recently, [16] used a combination of DenseNet201 and Grad-CAM visualization to not only classify pediatric pneumonia but also provide explainability by highlighting the affected lung regions, aiding clinical interpretability.

Several comparative studies have also evaluated the performance of multiple CNN models in pediatric pneumonia detection. For instance, [17] proposed a hybrid model combining CNN feature extraction with support vector machines (SVMs) and achieved notable performance improvements over standalone models. Additionally, [18] conducted an in-depth review of deep learning methods for pneumonia detection and noted that while transfer learning significantly enhances performance, model generalizability across populations and image acquisition settings remains a critical challenge.

Despite the promising results reported across these studies, most research has focused on a limited set of models or datasets, and comparative analyses using more recent architectures are relatively scarce. In response, our study explores and benchmarks three state-of-the-art CNNs—ResNetRS, RegNet, and EfficientNetV2—on pediatric chest X-ray images to assess their effectiveness in real-world diagnostic settings. By incorporating modern training procedures and robust evaluation metrics, this work contributes to the growing body of literature on AI-based pediatric pneumonia diagnosis and offers insights into the clinical applicability of advanced deep learning models.

# Method

**Dataset**

This study utilized a curated subset of a publicly available pediatric chest X-ray (CXR) dataset originally compiled by Kermany et al. (2018). The complete dataset contains 5,856 anterior-posterior chest radiographs of pediatric patients aged between one and five years, categorized into 3,883 pneumonia cases (including 2,538 bacteria and 1,345 viral) and 1,349 normal cases. For our experiments, we selected a representative subset of 1,000 images through stratified random sampling to preserve the original class distribution across pneumonia and normal categories. All images were visually inspected to exclude low-quality or unreadable scans. The Institutional Review Board (IRB)/Ethics Committee approvals were obtained for this dataset [1,19].

To ensure consistency and compatibility with pretrained convolutional neural network (CNN) models, all images were resized to 224 × 224 pixels. Preprocessing included normalization using the standard ImageNet mean and standard deviation values. To enhance generalization and reduce overfitting, we applied data augmentation techniques such as random horizontal flipping, small rotations (±10 degrees), zooming, and brightness adjustments. The dataset was split into 80% training and 20% testing sets, with stratification to maintain class balance. Five-fold cross-validation was used on the training set to evaluate model stability and performance across different data splits.

**Deep Learning Models**

In this study, we employed three state-of-the-art convolutional neural network (CNN) architectures—ResNetRS, RegNet, and EfficientNetV2—to perform binary classification of pediatric chest X-ray images into pneumonia or normal categories. All models were pretrained on the ImageNet dataset to leverage rich feature representations and were subsequently fine-tuned on our pediatric X-ray dataset using transfer learning.

*ResNetRS*

ResNetRS (ResNet-Ranked Scaled) [20] is a family of deep residual networks proposed by Google Research in 2021 as an improved variant of the traditional ResNet architecture. While standard ResNets use a fixed design, ResNetRS introduces several refinements aimed at enhancing training stability and accuracy. These include:

(a) Standardized convolutions and weight standardization, which normalize the weight distributions during training to stabilize gradients and improve convergence.

(b) Squeeze-and-Excitation (SE) blocks, which adaptively recalibrate channel-wise feature responses by modeling interdependencies between channels.

(c) Stochastic depth, a regularization technique that randomly drops residual connections during training, thereby improving model robustness and preventing overfitting.

(d) Swish activation function and adaptive scaling of width and depth across blocks.

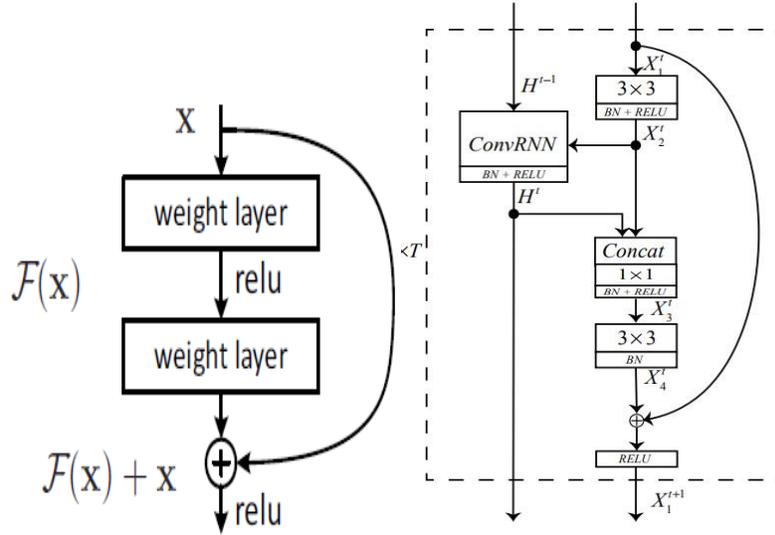

*Figure 2 Residual block in the original ResNet (left). The RegNet module (right) [20]*

We utilized the ResNetRS-50 variant, which has 50 layers and offers a strong balance between accuracy and computational efficiency. The final classification layer was replaced with a dense output layer containing a single neuron with sigmoid activation. Only the top convolutional block and batch normalization layers were unfrozen during fine-tuning, allowing the model to adapt high-level features to the pediatric pneumonia domain while preserving the generalized representations learned from ImageNet.

*RegNet*

RegNet (Regularized Network Design Space) [21], introduced by Facebook AI Research, is a family of CNNs derived from a simple design space where width, depth, and bottleneck ratios are parameterized using quantized linear functions. This structured approach leads to architectures that are computationally efficient and scalable without compromising accuracy. RegNet differs from earlier models by emphasizing design regularity over manual tuning or neural architecture search.

The specific variant used in our study is RegNetY-064, which incorporates:

(a) Group convolutions to reduce parameter redundancy while maintaining representational power.

(b) Bottleneck residual blocks that compress and expand features for efficient information flow.

(c) Squeeze-and-Excitation (SE) attention blocks to model channel-wise interdependencies and improve feature calibration.

(d) Efficient stage-wise block scaling, where both width and depth are increased progressively throughout the network.

Due to its modular and systematic architecture, RegNetY is well-suited for transfer learning tasks in medical imaging. The classification head was replaced with a dropout layer followed by a sigmoid-activated neuron for binary output. Full model fine-tuning was performed with a gradual unfreezing strategy across the training epochs.

*EfficientNetV2*

EfficientNetV2 [22], developed by Google Brain, is the successor to EfficientNet and is designed for both accuracy and speed. It achieves faster training with higher accuracy by incorporating two key innovations:

(a) Fused-MBConv layers, which replace the standard MBConv (Mobile Inverted Bottleneck Convolution) blocks in the early stages with regular convolution + batch normalization layers, resulting in faster convergence and reduced memory consumption.

(b) Progressive learning, a training approach that gradually increases the input image size and data augmentation strength throughout training. This helps stabilize early learning and reduce overfitting, particularly in small datasets.

EfficientNetV2 uses compound scaling to balance network depth, width, and resolution. We adopted the EfficientNetV2-Small (EffNetV2-S) variant due to its excellent performance on lightweight medical tasks and suitability for smaller datasets like pediatric chest X-rays. This model incorporates SE blocks and Swish activation functions, which enhance learning capacity and nonlinearity. We customized the classifier head to include batch normalization, dropout (p=0.3), and a final sigmoid activation for binary classification.

Unlike earlier EfficientNet models that required extensive tuning for small medical datasets, EfficientNetV2's improved learning dynamics make it particularly well-suited for our fine-tuning workflow. The entire model was unfrozen during training to enable comprehensive domain adaptation. These three models were selected due to their robust performance in general image classification, their proven transferability, and their ability to handle limited, class-imbalanced datasets—a common constraint in pediatric imaging tasks. Their architectural diversity also allowed us to compare CNNs with different building principles (residual, regularized, and mobile-efficient designs) under a uniform experimental setting.

**Experiment Setup**

All models were implemented in PyTorch and trained in an NVIDIA RTX GPU environment. The training configuration included the Adam optimizer with a learning rate of 1e-4, batch size of 32, and up to 30 epochs with early stopping based on validation loss. To further stabilize training, we employed gradient clipping with a maximum norm of 5.0 and used mixed-precision training via automatic mixed precision (AMP).

**Results**

We evaluated the performance of ResNetRS, RegNet, and EfficientNetV2 models on the held-out test set derived from our curated pediatric chest X-ray dataset. Each model was trained using 5-fold cross-validation and evaluated based on classification accuracy and sensitivity (recall)—two metrics of particular clinical importance in diagnostic settings.

*Table 1 Classification performance of the deep learning models*

| Model | Accuracy | Sensitivity |
|---|---|---|
| ResNetRS | 91.9 | 89.3 |
| RegNet | 92.4 | 90.1 |
| EfficientNetV2 | 88.5 | 88.1 |

Accuracy reflects overall classification correctness, while sensitivity quantifies the model's ability to correctly identify pneumonia cases, which is essential in avoiding missed diagnoses. The results are summarized in Table 1. Among the tested architectures, RegNet achieved the highest classification performance with an accuracy of 92.4% and sensitivity of 90.1%, indicating its strong ability to correctly identify pneumonia cases while maintaining overall reliability. ResNetRS followed closely, achieving an accuracy of 91.9% and sensitivity of 89.3%, which demonstrates consistent performance and generalization. EfficientNetV2 also performed well, with an accuracy of 88.5% and sensitivity of 88.1%, although slightly lower than the other two models.

Overall, the results indicate that modern CNN architecture pretrained on large-scale natural image datasets can be successfully adapted for pediatric medical imaging tasks. Among the tested models, RegNet emerges as the most promising candidate for real-world deployment in pneumonia screening tools, balancing predictive performance with architectural efficiency.

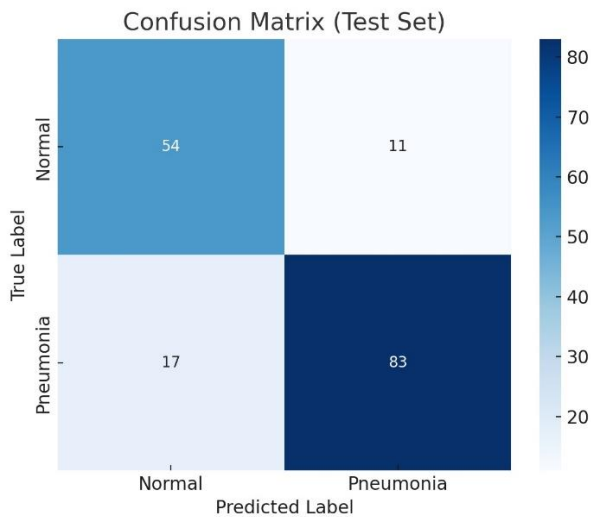

*Figure 3 Confusion matrix of RegNet model on testing data*

The confusion matrix presented in Figure 3 provides a detailed breakdown of the model's performance on the held-out test set. Out of the total pneumonia cases, the model correctly identified 83 true positives, while 17 cases were missed, resulting in false negatives. Among the normal cases, 54 were correctly classified as true negatives, and 11 were incorrectly predicted as pneumonia, leading to false positives. These results reflect a strong balance between sensitivity and specificity, with the model demonstrating a reliable ability to detect pneumonia while maintaining a low false positive rate. The relatively low number of false negatives is particularly important in clinical contexts, as it reduces the risk of untreated pneumonia progressing to severe complications. Overall, the confusion matrix confirms the model's robust diagnostic potential for pediatric pneumonia detection.

# Conclusion

In this study, we investigated the performance of three state-of-the-art convolutional neural network architectures—ResNetRS, RegNet, and EfficientNetV2—for the task of automated pediatric pneumonia classification using chest X-ray images. We curated and preprocessed a subset of 1,000 labeled images from a publicly available pediatric dataset and applied transfer learning techniques to adapt each model for binary classification. All models were trained and evaluated using a standardized pipeline involving five-fold cross-validation and an 80-20 train-test split. Our results demonstrated that all three models were capable of effectively distinguishing between pneumonia and normal chest radiographs in pediatric patients. Among them, RegNet achieved the highest accuracy (92.4%) and sensitivity (90.1%), followed closely by ResNetRS. EfficientNetV2 also showed competitive performance, although slightly lower than the other two. The confusion matrix further confirmed the models' ability to minimize false negatives and false positives, both of which are critical in clinical diagnostics.

Overall, this study highlighted the viability of modern CNN-based transfer learning approaches in supporting pediatric pneumonia diagnosis. The models we evaluated showed strong diagnostic performance and could serve as reliable tools in real-world settings, particularly in low-resource environments where expert radiological interpretation may be limited. Future work may focus on extending this analysis to larger and more diverse datasets, incorporating multi-class classification (e.g., distinguishing bacterial vs. viral pneumonia), and integrating model predictions into clinical decision-support systems.

# References

[1] Kermany DS, Goldbaum M, Cai W, Valentim CC, Liang H, Baxter SL, McKeown A, Yang G, Wu X, Yan F, Dong J. Identifying medical diagnoses and treatable diseases by image-based deep learning. cell. 2018 Feb 22;172(5):1122-31.

[2] Irede EL, Aworinde OR, Lekan OK, Amienghemhen OD, Okonkwo TP, Onivefu AP, Ifijen IH. Medical imaging: a critical review on X-ray imaging for the detection of infection. Biomedical Materials & Devices. 2024 Jul 15:1-45.

[3] Çallı E, Sogancioglu E, Van Ginneken B, van Leeuwen KG, Murphy K. Deep learning for chest X-ray analysis: A survey. Medical image analysis. 2021 Aug 1;72:102125.

[4] Sharifi S, Donyadadi A. Detection and diagnosis of congenital heart disease from chest X-rays with deep learning models. International Journal of Applied Data Science in Engineering and Health. 2025 Jan 2;1(1):1-9.

[5] Kohli PS, Arora S. Application of machine learning in disease prediction. In2018 4th International conference on computing communication and automation (ICCCA) 2018 Dec 14 (pp. 1-4). IEEE.

[6] Norouzi F, Machado BL. Predicting Mental Health Outcomes: A Machine Learning Approach to Depression, Anxiety, and Stress. International Journal of Applied Data Science in Engineering and Health. 2024 Oct 31;1(2):98-104.

[7] Uddin S, Khan A, Hossain ME, Moni MA. Comparing different supervised machine learning algorithms for disease prediction. BMC medical informatics and decision making. 2019 Dec;19(1):1-6.

[8] Ghasemi F, Sharifi S. Heart Failure Prediction Using Support Vector Machine. International Journal of Novel Research in Life Sciences. 2025 Feb 1.

[9] Anantharajan S, Gunasekaran S, Subramanian T. MRI brain tumor detection using deep learning and machine learning approaches. Measurement: Sensors. 2024 Feb 1;31:101026.

[10] Whig P, Shadadi E, Kouser S, Alamer L. Machine learning approaches for early detection and management of musculoskeletal conditions. International Journal of Computational Vision and Robotics. 2025;15(1):104-17.

[11] Wang L. Mammography with deep learning for breast cancer detection. Frontiers in oncology. 2024 Feb 12;14:1281922.

[12] Abbasi H, Afrazeh F, Ghasemi Y, Ghasemi F. A shallow review of artificial intelligence applications in brain disease: stroke, Alzheimer's, and aneurysm. International Journal of Applied Data Science in Engineering and Health. 2024 Oct 5;1(2):32-43.

[13] Rajpurkar P, Irvin J, Zhu K, Yang B, Mehta H, Duan T, Ding D, Bagul A, Langlotz C, Shpanskaya K, Lungren MP. Chexnet: Radiologist-level pneumonia detection on chest x-rays with deep learning. arXiv preprint arXiv:1711.05225. 2017 Nov 14.

[14] Stephen O, Sain M, Maduh UJ, Jeong DU. An efficient deep learning approach to pneumonia classification in healthcare. Journal of healthcare engineering. 2019;2019(1):4180949.

[15] Abiyev RH, Ma'aitaH MK. Deep convolutional neural networks for chest diseases detection. Journal of healthcare engineering. 2018;2018(1):4168538.

[16] Ali SA, Kumar MS. Detecting pneumonia from X-ray images using transfer learning. In2020 IEEE International Conference on Advent Trends in Multidisciplinary Research and Innovation (ICATMRI) 2020 Dec 30 (pp. 1-4). IEEE.

[17] Toğaçar M, Ergen B, Cömert Z. COVID-19 detection using deep learning models to exploit Social Mimic Optimization and structured chest X-ray images using fuzzy color and stacking approaches. Computers in biology and medicine. 2020 Jun 1;121:103805.

[18] Chowdhury ME, Rahman T, Khandakar A, Mazhar R, Kadir MA, Mahbub ZB, Islam KR, Khan MS, Iqbal A, Al Emadi N, Reaz MB. Can AI help in screening viral and COVID-19 pneumonia?. Ieee Access. 2020 Jul 20;8:132665-76.

[19] Kermany D, Zhang K, Goldbaum M. Labeled optical coherence tomography (oct) and chest x-ray images for classification (2018). Mendeley Data, v2 https://doi. org/10.17632/rscbjbr9sj https://nihcc. app. box. com/v/ChestXray-NIHCC.

[20] Bello I, Fedus W, Du X, Cubuk ED, Srinivas A, Lin TY, Shlens J, Zoph B. Revisiting resnets: Improved training and scaling strategies. Advances in Neural Information Processing Systems. 2021 Dec 6;34:22614-27.

[21] Radosavovic I, Kosaraju RP, Girshick R, He K, Dollár P. Designing network design spaces. InProceedings of the IEEE/CVF conference on computer vision and pattern recognition 2020 (pp. 10428-10436).

[22] Tan M, Le Q. Efficientnetv2: Smaller models and faster training. InInternational conference on machine learning 2021 Jul 1 (pp. 10096-10106). PMLR.